\documentstyle[12pt,psfig]{article}

\begin{document}
\newcommand{\be}{\begin{equation}}
\newcommand{\ee}{\end{equation}}
\newcommand{\bea}{\begin{eqnarray}}
\newcommand{\eea}{\end{eqnarray}}
\renewcommand{\topfraction}{1.}

\title{On the Universality of Nonperturbative effects\\
            in Stabilized 2D Quantum Gravity}

\author{{\bf Oscar Diego}\thanks{e-mail: imtod67@cc.csic.es} \\
        {\em Instituto de Estructura de la Materia, CSIC } \\
        {\em Serrano 123, 28006 Madrid} \\
        {\em Spain } }

\date{\mbox{ }}

\maketitle

\thispagestyle{empty}

\begin{abstract}
In this letter I study the universality of the
nonperturbative effects and the vacua structure of the
stochastic stabilization of the matrix models which defines
Pure 2D Quantum Gravity. I show also that there is not
tunneling, in the continuum limit, between the one-arc and
three-arc solutions of the simplest matrix model which
defines the flow between Pure Gravity and the Lee-Yang
model.

\end{abstract}

\begin{flushright}
\vspace{-13.5 cm} {IEM-FT-94/94}
\end{flushright}
\baselineskip=21pt
\vfill
\newpage
\setcounter{page}{1}

{\bf 1.} Pure Quantum Gravity in two dimensions is an ill
defined theory because the topological expansion is given
by a non Borel summable series\cite{MATRIX,BOREL,ZINN}.
In the matrix models approach\cite{MATRIX}, the topological
expansion is defined by the perturbative series of a matrix
model with an unbounded potential, and the non Borel summability
of the topological expansion is related to tunneling from the
local minimun of the matrix potential to the unbounded
region\cite{ZINN}.

There are several proposals for stabilize an unbounded
matrix model\cite{GSTA}.
In the stochastic stabilization\cite{STAB} the matrix model is
mapped to a one dimensional matrix model, where the unbounded
region of the potential is related to the local minimun of the
stabilized potential.
In fact, the perturbative ground state of the one dimensional
stabilized model is the original matrix model.

In a previous paper\cite{OSC2} the analytical
ground state of the stochastic
stabilization of the fourth matrix potential has been studied,
taking into account nonperturbative corrections. The
nonperturbative effects reproduce the asymptotic behaviour of
the topological expansion, and is related to tunneling from
the perturbative vacuum, which defines the topological
expansion, to a nonperturbative vacuum, which break down the
symmetries of the original matrix model.

The stabilized matrix model is given by a one dimensional Fermi
gas. In the planar limit the Fermi energy is placed at the local
minimun of the stabilized potential\cite{STAB}. The perturbative
corrections decrease the Fermi energy, and it become below the
local minimum\cite{OSC2,MIG,OSC1}. This is very important
because the nonperturbative ground state is given by a
combination
\begin{equation}
\Psi = \Psi_{pert} + \Psi_{nonpert}
\label{eq:1}
\end{equation}
where $\Psi_{pert}$ is a perturbative  state around the global
minimun of the stabilized potential which defines the
topological expansion. Around the local minimun $\Psi_{nonpert}$
is the state with energy below the potential, hence it cannot be
a perturbative eigenstate around the local minimun because the
WKB wave functions around the local minimum have energy above
it. Henceforth $\Psi_{nonpert}$ can be interpreted as a
nonperturbative vacuum and (\ref{eq:1}) defines a ground state
where the nonperturbative effects are given by tunneling between
a perturbative vacuum and a nonperturbative one\cite{OSC2}.

In this short letter I study the universality of the
vacua structure and the nonperturbative
effects in stabilized matrix models. The
cubic, fourth and sixth matrix models are considered. I show
that there is not tunneling between one-arc and three-arc
solution in the sixth matrix model.

{\bf 2.} The cubic potential is the simplest matrix
model which defines Pure Quantum Gravity:
\begin{equation}
W = Tr \Phi^{2} - \frac{2}{3} g Tr \phi^{3}
\end{equation}
and the stabilized potential is\cite{CUBIC}
\be
V = \frac{1}{2} \left \{ g^{2} \lambda^{4} - 2 g \lambda^{3} +
\lambda^{2} + 2 g \lambda - 1 \right \}.
\ee

The perturbative Fermi energy is given by the condition:
\be
\frac{N}{\pi} \int d \lambda \sqrt{2 ( E_{F} - V ) } = N -
\frac{1}{2}
\ee
where the minus sign in $N - 1/2$
is because the Fermi gas is given by $N$ free fermions
which must fill
the first $N$ eigenvalues of the WKB quantization condition
\be
\frac{N}{\pi} \int d \lambda \sqrt{2(E_{F} - V)} = n + \frac{1}{2}
\ee
from $n=0$ to $n=N-1$.

The difference between the Fermi energy and the value of the
potential at the local minimun $V_{min}$:
\begin{equation}
\Delta = E_{F} - V_{min} = \Delta_{0} + \frac{1}{N} \Delta_{1} +
\cdots
\label{eq:2}
\end{equation}
is zero at leading order\cite{STAB} and negative at subleading
order\cite{OSC1}:
\bea
\Delta_{0} & = & 0 \\
\Delta_{1} & = & - \frac{1}{2} \left [ \frac{1}{\pi} \int d
\lambda \frac{1}{\sqrt{2(E_{F} - V)}} \right ]^{-1} .
\eea

The fourth potential is
\begin{equation}
W = Tr \Phi^{2} - \frac{2}{4} g Tr \phi^{4}
\end{equation}
and its stabilized potential in the Hartree-Fock
approach\footnote{ The stabilized potential of the fourth
and next order matrix models has interaction terms, but the
Hartree-Fock approach is exact to all orders in $1/N$.}
is\cite{CUARTO}
\begin{equation}
V = \frac{1}{2} \left \{ g^{2} \lambda^{6} - 2 g \lambda^{4} +
\lambda^{2} + 2 g \lambda^{2} - 1 \right \} + \frac{1}{N}
\frac{1}{2} g \lambda^{2} + \mbox{Fock}.
\end{equation}

The perturbative Fermi energy is given now by:
\be
\frac{N}{\pi} \int_{-a}^{a} d \lambda \sqrt{2(E_{F} - V) } -
\frac{1}{\pi} \frac{1}{2} g \int_{a}^{b} d \lambda
\frac{\lambda^{2}}{\sqrt{2(E_{F} - V)}} + O(1/N) = N -
\frac{1}{2}.
\ee
$\Delta$ is zero at leading order and negative at subleading
order\cite{OSC2}
\be
\Delta_{1} = - b g \sqrt{b^{2} - a^{2}}
\ee
where $a$ is the cut of the eigenvalue density and $b$ is the
local minimun of the stabilized potential\cite{CUARTO}.

{\bf 3.} The sixth potential:
\be
V = Tr \Phi^{2} - \frac{2}{4} g Tr \phi^{4} +
\frac{2}{6} \alpha Tr \Phi^{6}
\ee
is the simplest potential which defines the flow between Pure
Gravity and the Lee-Yang model\cite{MATRIX,SHENKER}. If
$\alpha$ is positive there are three kind of vacua in the planar
limit\cite{ARCS}: the one-arc vacuum, where the eigenvalue
density $\rho(\lambda)$ is defined on one interval; the two-arc
vacuum, where $\rho(\lambda)$ is defined on two intervals and an
infinite family of three-arc vacua where $\rho(\lambda)$ is
defined on three intervals.

Pure Quantum Gravity is the continuum limit of the one-arc
vacuum. In this model there is a line of critical points which
defines Pure Quantum Gravity. The line ends at the
tricritical point which defines
the Lee-Yang model. But at the critical
line there are three-arc and two-arc
solutions. In Refs.~\cite{SHENKER,ARCS} it has been argued
that the tunneling between one-arc solution and
three-arc solutions at the critical line is the origin of the
nonperturbative instability of
the KdV flow between Lee-Yang an Pure Gravity.
I will show that
the tunneling between one-arc and two or three arc solutions
does not survive in the continuum limit.

The stabilized potential in the Hartree-Fock approach
is\cite{OSC0}
\bea
V & = & V_{0} + \frac{1}{N} V_{1} + \mbox{Fock} \nonumber \\
V_{0} & = & \frac{1}{2} \left \{ \alpha^{2} \lambda^{10}
- 2 g \alpha \lambda^{8}
+ ( g^{2} + 2 \alpha ) \lambda^{6}
- 2 ( g + \alpha ) \lambda^{4} \right \} \nonumber \\
  & + & \frac{1}{2} \left \{ ( 1 + 2 g - 2 \alpha \omega )
\lambda^{2} - 1 \right \} \nonumber \\
V_{1} & = & \frac{1}{N}
\left \{ \frac{1}{2} g \lambda^{2} - \frac{3}{2} \alpha
\lambda^{4} \right \}
\eea
where $\omega$ is a self-consistent parameter
\bea
\omega & = & \frac{1}{N} \langle Tr \Phi^{2}
             \rangle \nonumber \\
       & = & \frac{1}{\pi} \int d \lambda
             \lambda^{2} \sqrt{2 (E_{F} - V)}
   	     + O(1/N^{2}) .
\eea

In the planar limit there is an infinite set of ground states of
the above Hamiltonian which are the multi-cut solutions of the
original matrix model\cite{OSC0}.

The one-arc and three-arc solutions are represented in the
Figure~1. The one-arc solution has four local degenerate
minima $b$, $-b$, $c$ and $-c$, and the Fermi energy is placed
at these local minima. Hence, the eigenvalues are restricted to
the interval $(-a,a)$ where $a$ is the cut of the eigenvalue
density of the original matrix model.

The three-arc solutions have four nondegenerate
local minima and the Fermi energy is placed at the minimun $b$.
There are eigenvalues in the three intervals $(-d,-c)$, $(-a,a)$
and $(c,d)$ of figure~1. These intervals are the support of the
eigenvalue density of the three-arc solutions of the original
matrix model.

\begin{figure}[t]
\centerline{ \vbox{
\hbox{\psfig{figure=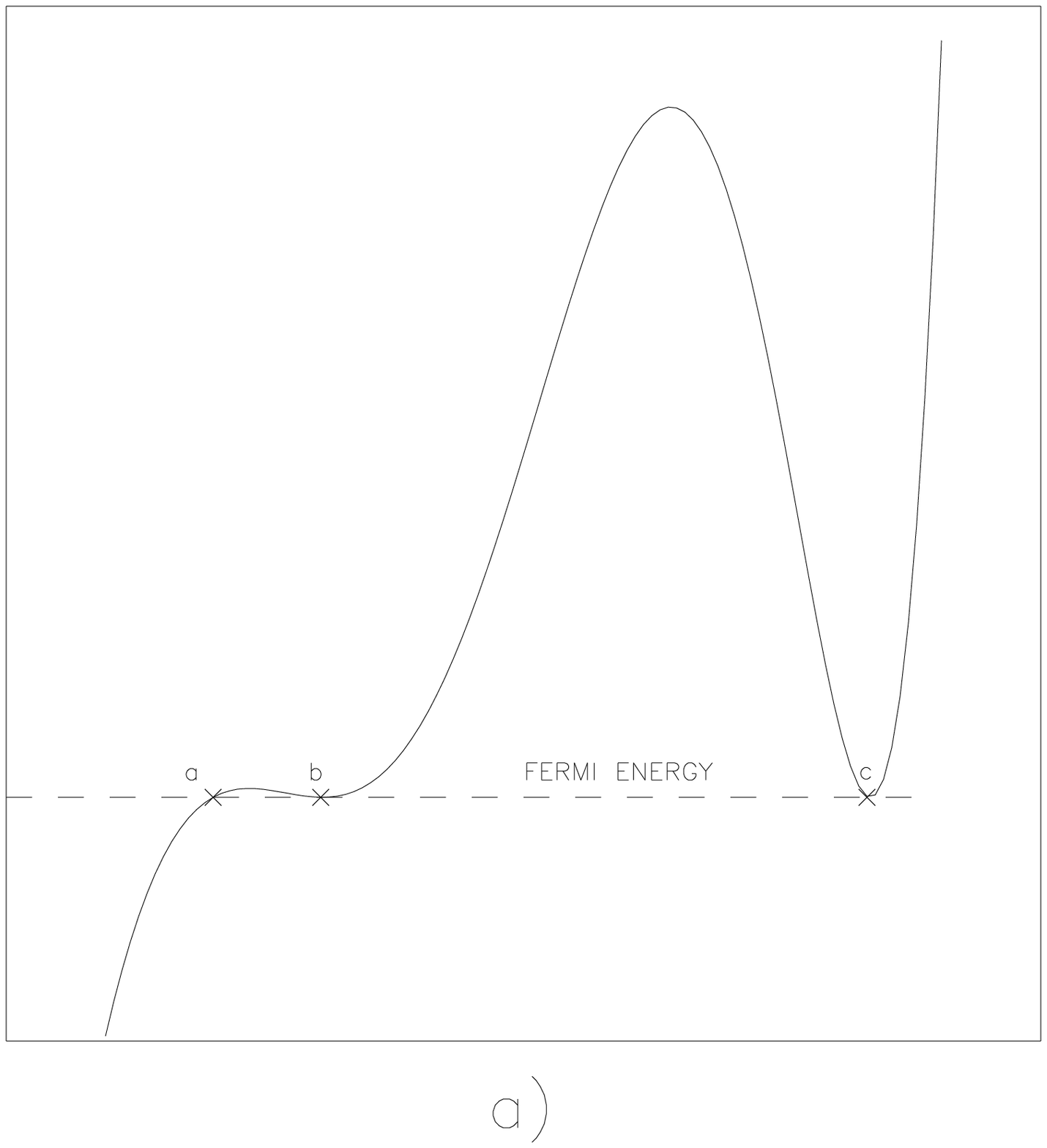,height=6.cm,width=6.cm,bbllx=4.cm,bblly=1.cm,bburx=20.5cm,bbury=18cm}
\psfig{figure=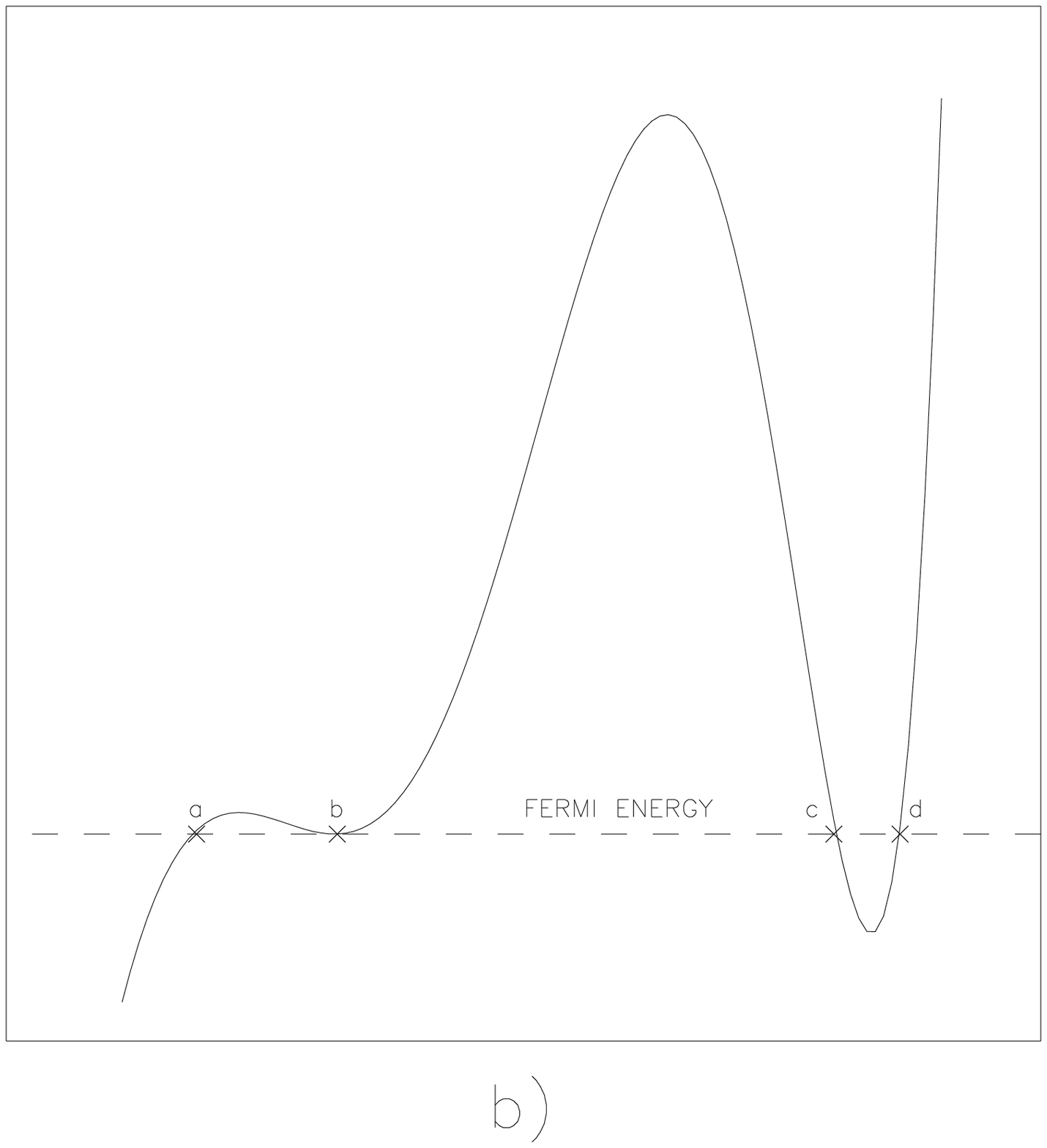,height=6.cm,width=6.cm,bbllx=4.cm,bblly=1.cm,bburx=20.5cm,bbury=18cm}}
}}
\caption{\tenrm\baselineskip=12pt
Stabilized potential for $\lambda$ positive (the potential
is even in $\lambda$) in the Hartree approach for the
sixth potential: a) One-arc solution, b) Three-arc solution.}
\end{figure}

In the one-arc solution the Fermi energy is given by the
condition
\be
\frac{N}{\pi} \int_{-a}^{a} d \lambda \sqrt{2(E_{F} - V) }
- \frac{1}{\pi} \int_{a}^{b} d \lambda
\frac{V_{1}}{\sqrt{2(E_{F} - V)}} + O(1/N) = N - \frac{1}{2}
\ee
which must be supplemented by a condition to the self-consistent
parameter $\omega$ at leading order
\be
\omega = \frac{1}{\pi} \int_{-a}^{a} d \lambda
\lambda^{2} \sqrt{2 (E_{F} - V)}.
\ee

In the one-arc solution (figure~1.a) the difference between the
Fermi energy and the value of the potential at the two local
minima are zero at leading order and at subleading order are:
\bea
\Delta_{1} (b) & = & \left [ - 2 \alpha ( c^{2} - b^{2} ) +
\frac{1}{c \sqrt{c^{2} - a^{2}}} \left \{
V_{1} ( b ) - V_{1} ( c ) \right \} \right ] \nonumber \\
 & \times & \left [ \frac{1}{b \sqrt{b^{2} - a^{2}}}
- \frac{1}{c \sqrt{c^{2} - a^{2}}} \right ]^{-1} \nonumber \\
\Delta_{1} (c) & = & \left [ - 2 \alpha ( c^{2} - b^{2} ) +
\frac{1}{b \sqrt{b^{2} - a^{2}}} \left \{
V_{1} ( b ) - V_{1} ( c ) \right \} \right ] \nonumber \\
 & \times & \left [ \frac{1}{b \sqrt{b^{2} - a^{2}}}
- \frac{1}{c \sqrt{c^{2} - a^{2}}} \right ]^{-1}.
\eea
Numerically one can check that $\Delta_{1} ( b )$ is negative
and $\Delta_{1} ( c ) $ is positive. Hence, at subleading order
in $1/N$ the Fermi energy is below the local minimun $b$ and
above the local minimun $c$.

This result is very natural because the local minimun $c$ is
related to the nonzero minimun of the original potential. Well
defined configurations of the original potential must be
related to perturbative ground states of the stochastic
stabilization\cite{GREEN}.

Henceforth there exist tunneling between the one-arc solution,
where all the eigenvalues are restricted to the central well of
the stabilized potential and the perturbative vacuum with an
eigenvalue around the local minimun $c$. This perturbative
vacuum is a three-arc vacuum. Hence the tunneling from the main
well to the local minimun $c$ can be interpreted as tunneling
between the one-arc vacuum and the three-arc vacuum. From
standard WKB calculation the tunneling is proportional to:
\be
\exp{ \left \{ - N ( \Gamma_{1} + \Gamma_{2} ) \right \} }
\ee
where
\bea
\Gamma_{1} & = & \int_{a}^{b} d \lambda \sqrt{ 2 ( V - E_{F} ) }
\nonumber \\
\Gamma_{2} & = & \int_{b}^{c} d \lambda \sqrt{ 2 ( V - E_{F} )}.
\eea

The critical point $g_{c}$ which defines perturbative Pure
Quantum Gravity, is given when $a=b$ in the one-arc solution
(figure~1.a). The double scaling limit\cite{MATRIX} is the
limit:
\bea
N & \rightarrow & \infty \nonumber \\
g & \rightarrow & g_{c} \nonumber \\
N ( g - g_{c} )^{5/4} & = & z
\eea
where $z$ is finite.
In the double scaling limit $\Gamma_{1}$ goes to zero and
\be
\exp{ \left \{ - N \Gamma_{1} \right \} } \longrightarrow
\exp{\{- C z \}}
\label{eq:3}
\ee
where $C$ is a universal
constant for all matrix models\cite{MIRAMONTES}.
$\Gamma_{2}$ remains finite at $g = g_{c}$, and, in
the double scaling limit
\be
\exp{ \left \{ - N \Gamma_{2} \right \} } \longrightarrow 0
\ee
hence, in the double scaling limit the tunneling between the
perturbative one-arc solution and the perturbative three-arc
solution does not survive. In the double scaling limit the
nonperturbative effects are given only by $\Gamma_{1}$ which
can be interpreted as tunneling between the main well of the
stabilized potential and the local minimun $b$. Because the
Fermi energy is below the value of the potential at the
local minimun $b$, the stabilized potential of the sixth matrix
model has the same nonperturbative behaviour that the model
studied in Ref.~\cite{OSC2}, and following the same reasoning
that in this reference one can construct the ground state as a
linear combination of a perturbative ground state around the
global minimun and a wave function around the local minimun $b$,
which is nonperturbative.

Henceforth, the origin of the instability of the flow between
Lee-Yang and Pure Gravity is the tunneling between the
perturbative vacuum which defines Pure Gravity and a
nonperturbative vacuum which break down the symmetries of the
matrix model. This is also the origin of the instability of
the loop equation\cite{OSC2,OSC1}.

In the two-arc vacuum the distribution of eigenvalues
differs macroscopically from the
one-arc vacuum, but only tunneling of one eigenvalue may survive
in the double scaling limit\cite{DAVID}. Hence there is not
tunneling between one-arc and two-arcs vacuum in the double
scaling limit.

\vfill

\pagebreak


\begin{thebibliography}{99}

\bibitem{MATRIX} E. Br\'ezin and V. A. Kazakov,
{\em Phys. Lett.} {\bf B236}, 144 (1990);
M. Douglas and S. Shenker, {\em Nucl. Phys.} {\bf B335}, 635
(1990); D. J. Gross and A. A. Migdal,  {\em Phys. Rev. Lett.}
{\bf 64}, 717 (1990); {\em Nucl. Phys.} {\bf B340}, 333 (1990).

\bibitem{BOREL} S. Shenker, {\em The strenght of
nonperturbative effects
in string theory}, Rutgers Preprint RU-90-47,
in Cargese Workshop on Random
Surfaces, Quantum Gravity and Strings, Cargese 1990.

\bibitem{ZINN} P. Ginsparg and J. Zinn-Justin, {\em Phys. Lett.}
{\bf B255}, 189 (1991); {\em Action principle and large order
behaviour of nonperturbative
gravity} in Random Surfaces,
Quantum Gravity and Strings, (Cargese 1990);
 B. Eynard and J. Zinn-Justin, {\em Phys. Lett.} {\bf B302},
396(1993).

\bibitem{GSTA} F. David, {\em Mod. Phys. Lett.}
{\bf A5}, 1651 (1990); P. G. Sylvestrov and A. S. Yelkovsky,
{\em Phys. Lett.} {\bf B251}, 525 (1990);
F. David, {\em Nucl. Phys.} {\bf B348}, 507 (1991);
M. Saadi and G. Zemba, {\em Int. Jour. Mod. Phys.}{\bf A7}, 501 (1992);
T. R. Morris, {\em Nucl. Phys.} {\bf B356}, 703 (1991);
S. Dalley, C. Johnson and T. Morris, {\em Nucl. Phys.} {\bf
B368}, 625 (1992); ibid. 655; {\em Phys. Lett.} {\bf B291},
11 (1992); {\em Nucl. Phys.} (Proc. Suppl) {\bf 25A}, 87 (1992);
T. Hollowood, J. L. Miramontes, A. Pasquinucci and C. Nappi,
{\em Nucl. Phys.} {\bf B373}, 247(1992).

\bibitem{STAB} J. Greensite and M. Halpern; {\em Nucl. Phys.}
{\bf B242}, 167(1984); E. Marinari and G. Parisi,
{\em Phys. Lett.} {\bf B274}, 537 (1990);
M. Karliner and A. Migdal, {\em Mod. Phys. Lett.}
{\bf A5}, 2565 (1990); J. Ambj\o rn and G. Greensite and
S. Varsted, {\em Phys. Lett} {\bf B249}, 411 (1990);
J. L. Miramontes, J. S. Guill\'en and M. A. H. Vozmediano,
{\em Phys. Lett.} {\bf B253}, 38 (1991); A. Dabholkar,
{\em Nucl. Phys.} {\bf B368}, 293 (1992).

\bibitem{OSC2} O. Diego, {\em Nonperturbative Stochastic
definitions of 2D Quantum Gravity}, to appear in
{\em Mod. Phys.Lett. A}, {\em Bulletin board } hepth/9310079.

\bibitem{MIG} M. Karliner, A. Migdal and B. Rusakov,
{\em Nucl. Phys.} {\bf B399},
514 (1993).

\bibitem{OSC1} O. Diego and J. Gonz\'alez, {\em Mod. Phys. Lett.}
{\bf A9}, 2253 (1994).

\bibitem{CUBIC} J. Ambj\o rn and G. Greensite, {\em Phys. Lett.},
{\bf B254} 66 (1991); J. Ambj\o rn, C. V. Johnson and
T. R. Morris, {\em Nucl. Phys.} {\bf B374}, 496 (1992);
J. Ambj\o rn and C. F. Kristjansen, {\em Int. J. Mod. Phys.}
{\bf A8}, 1259 (1993).

\bibitem{CUARTO} J. Gonz\'alez and M. A. H. Vozmediano,
{\em Phys. Lett.} {\bf B258}, 55 (1991).

\bibitem{SHENKER} T. Banks, M. Douglas, N. Seiberg and S. Shenker,
{\em Phys. Lett.} {\bf B238}, 279 (1990); M. Douglas, N. Sieberg
and S. Shenker, {\em Phys. Lett.} {\bf B244}, 381 (1990).

\bibitem{ARCS} J. Jurkiewicz, {\em Phys. Lett.} {\bf B245},
178 (1990); G. M. Cicta, L. Molinori and E. Montaldi,
{\em J. Phys.} {A23}, L421 (1990); G. Bhanot, G. Mandal and
O. Narayan, {\em Phys. Lett.} {\bf B251}, 388 (1990); K. Demeterfi,
N. Deo, S. Jain and C. Tan, {\em Phys. Rev.} {D 42}, 4105 (1990);
M. Sasaki and H. Suzuki, {\em Phys. Rev.} {\bf D43}, 4015 (1991).

\bibitem{OSC0} O. Diego and J. Gonz\'alez, {\em Mod. Phys. Lett.}
{\bf A7}, 3465 (1992).

\bibitem{GREEN} M. Claudson and M. Halpern, {\em Ann. Phys. (NY)}
166 (1986)33; {\em Phys. Lett.} {B151}, 281 (1985);
J. Greensite, {\em Nucl. Phys.} {\bf B390}, 439 (1993).

\bibitem{MIRAMONTES} J. L. Miramontes and J. S. Guill\'en,
{\em Nucl. Phys.} (Proc. Suppl.) {\bf 25A}, 195 (1992);
{\em Int. J. Mod. Phys.} {\bf A7}, 6457 (1992).

\bibitem{DAVID} F. David {\it Simplicial Quantum Gravity and
Random Lattices.} Lectures given at Les Houches 1992.
Saclay Preprint T93/028.

\end{thebibliography}
\end{document}